\documentclass{icsm}          
\usepackage{epsf}

\slacs{.4ex}

\renewcommand{\S}{{\cal S}}

\begin{document}
\title{Saturation and non-linear effects in diffractive processes}
\authori{O.V. Selyugin\footnote{Visiting Fellow of the Fonds
National pour la Recherche Scientifique (Belgium), on leave from
BLTPh, JINR, Dubna, Russia.} and J.R. Cudell}
\addressi{Institut de Physique, B\^at. B5a, Universit\'e
de Li\`ege\\4000 Li\`ege, Belgique}
\authorii{}     \addressii{}
\authoriii{}    \addressiii{}
\authoriv{}     \addressiv{}
\authorv{}      \addressv{}
\authorvi{}     \addressvi{}
\headauthor{O.V. Selyugin et al.}
\headtitle{Saturation and non-linear effects in diffractive
processes}
\lastevenhead{O.V. Selyugin et al.: Saturation and non-linear
effects in diffractive processes}

\maketitle

\begin{abstract}
Through a direct implementation of the saturation regime resulting
 from the unitarity limit in the impact parameter representation,
we explore various possibilities for the energy dependence of
hadronic scattering. We show that it is possible to obtain a good
description of the scattering amplitude from a hard pomeron
provided one includes non-linear effects.
\end{abstract}

\section{Introduction}

The most important results on the energy dependence of diffractive
hadronic scattering were obtained from first principles
(analyticity, unitarity and Lorentz invariance), which lead to
specific analytic forms for the scattering amplitude as a function
of its kinematical parameters --- $s$, $t$, and $u$. Analytic
$S$--matrix theory relates the high-energy behaviour of hadronic
scattering to the singularities of the scattering amplitude in the
complex angular momentum plane. One of the important theorems is
the Froissart--Martin bound \cite{frois} which states that the
high--energy cross section for the scattering of hadrons is
limited by \be \sigma_{\mathrm{tot}}^{\max}=
\frac{2\pi}{\mu^2}\log^2\left(\frac{s}{s_0}\right), \label{FM} \ee
where $s_0$ is a scale factor and $\mu$ the lightest hadron mass
({\it i.e.} the pion mass). As the coefficient in front of the
logarithm is very large, ``saturation of the Froissart--Martin
bound'' usually refers to an energy dependence of the total  cross
section rising as $\log^2s$ rather than to a total cross section
equal to (\ref{FM}).

Experimental data reveal that total cross sections grow with
energy. This means that the leading contribution in the
high--energy limit is given by the rightmost singularity in the
complex--$j$ plane, the pomeron, with intercept exceeding unity.
In the framework of perturbative QCD, the intercept is also
expected to exceed unity by an amount proportional to $\alpha_s$
\cite{lipatov1}. At leading--log, one obtains a leading
singularity at $J-1=12\log 2({\alpha_s }/{\pi})$. In this case,
the Froissart--Martin bound is soon violated.

But this is not the whole story, as there is another important
consequence of unitarity, which is connected with the
impact--parameter representation.

\section{Unitarity bound}

Unitarity of the scattering matrix $SS^+=1$ is connected with the
properties of the scattering amplitude in the impact parameter
representation as it is equivalent at high energy to a
decomposition in partial--wave amplitudes. As energy increases,
the scattering amplitude in impact parameter can saturate the
unitarity bound for some impact parameter $b_s$.

To satisfy the unitarity condition, there are different
prescriptions. Two of them are based on particular solutions of
the unitarity equation.

For two--particle elastic scattering, the latter can be written
\be \Im\bigl<p_1,p_2,\mathrm{out}|T|p_1,p_2,\mathrm{in}\bigr>=
\frac{(2\pi)^4}2\sum_{\gamma}\int\D\gamma\,\delta
\left(\sum_{r=1}^{2}p_r-\sum_{r=1}^{n}q_r\right)
|T_{\gamma\alpha}|^2. \ee The scattering amplitude in the impact
parameter representation is defined as \be
T(s,t)=\int_0^{\infty}b\,\D b J_{0}(b\Delta)f(b,s)\,. \ee with
\be \Im f(b,s)\leq1\,. \ee and \be \Im f(s,b)=\bigl[\Im
f(s,b)\bigr]^2+\bigl[\Re f(s,b)\bigr]^2. \ee

One of the possibilities is obtained in the $U$--matrix approach
\cite{trosh}: \be f(s,b)=\frac{U(s,b)}{1-\I U(s,b)}\,.
\label{U-matrix} \ee This solution leads to a nonstandard
behaviour of the ratio \cite{trosh} \be
\frac{\sigma_{\mathrm{el}}}{\sigma_{\mathrm{tot}}}\ \rightarrow\
1\,, \ee as $s\rightarrow\infty$. For the highest energies reached
so far ($\sqrt{s}=2.0$\,TeV), such a ratio is $0.25$. It is not
too far from the standard value $1/2$, but it is very far from the
solution in the $U$--matrix representation.

The second possible solution of the unitarity condition
corresponds to the eikonal representation \be
T(s,t)=\int_0^{\infty}b\,\D b J_0(b\Delta)
\bigl(1-\exp(-\chi(s,b)\bigr)\,. \ee with $t=\Delta^2$. If one
takes the eikonal phase in factorized form \be
\chi(s,b)=h(s)\,f(b)\,, \ee one usually supposes that, despite the
fact that the energy dependence of $h(s)$ can be a power \be
h(s)\sim s^{\Delta}\,, \ee the total cross section will satisfy
the Froissart bound \be \sigma_{\mathrm{tot}}\leq a\log^2s\,. \ee
We find in fact that the energy dependence of the imaginary part
of the amplitude and hence of the total cross section depends on
the form of $f(b)$, {\it i.e.} on the $s$ and $t$ dependence of
the slope of the elastic scattering amplitude. If $f(b)$ decreases
as a power of $b$, the Froissart--Martin bound will always be
violated. In the case of other forms of the $b$ dependence, a
special analysis \cite{dif04} is required. Note that the eikonal
form does not correspond to a saturation of the amplitude: in this
case, $\Im f(s,b)$ reaches the black disk limit only
asymptotically. Hence, the saturation of the black disk limit and
the eikonal representation lead to different results for the
scattering amplitude in the momentum transfer representation.

Let us take, as an  example, the hard plus soft pomeron model
\cite{DoLa,mrt} which includes two simple poles (a soft and a hard
pomeron) to describe $pp$ and $\bar p p$ scattering. In this case,
the $pp$--elastic scattering amplitude is proportional to the
hadron form factors and can be approximated at small $t$ by: \be
\Im T^0(s,t)\approx\left[h_{1}(s/s_0)^{\epsilon_1}
\E^{\alpha^{\prime}_1t\log(s/s_0)}+h_2(s/s_0)^{\epsilon_2}
\E^{\alpha^{\prime}_2t\log(s/s_0)}\right]F^2(t)\,. \label{ampl}
\ee where $h_1=4.7$  and $h_2 = 0.005$ are the coupling of the
``soft'' and ``hard'' pomerons, and $\epsilon_1 =0.0072$,
$\alpha^{\prime}_1=0.25$, and $\epsilon_2=0.45$,
$\alpha^{\prime}_2=0.20$ are the intercepts and the slopes of the
two pomeron trajectories. The normalization $s_0$ will be dropped
below and  $s$ contains implicitly the phase factor
$\exp(-\I\pi/2)$. $F^2(t)$ is  the square of the Dirac elastic
form factor, which can be approximated by the sum of three
exponentials \cite{book}. We then obtain in the impact parameter
representation a specific form for the profile function
$\Gamma(b,s)$ \cite{dif04}, which we show in Fig.~1. One can see
that at some energy and at small $b$, $\Gamma(b,s)$ reaches  the
black disk limit. For one--pomeron exchanges, this will be in the
region $\sqrt{s}\approx 1.5$\,TeV. If one adds to the model
2--pomeron exchanges, the resulting $\Gamma_2$ will saturate at
$\sqrt{s}=4.5$\,TeV.

\bfg[t] \vskip-15mm\bc \epsfysize=80mm
\centerline{\epsfbox{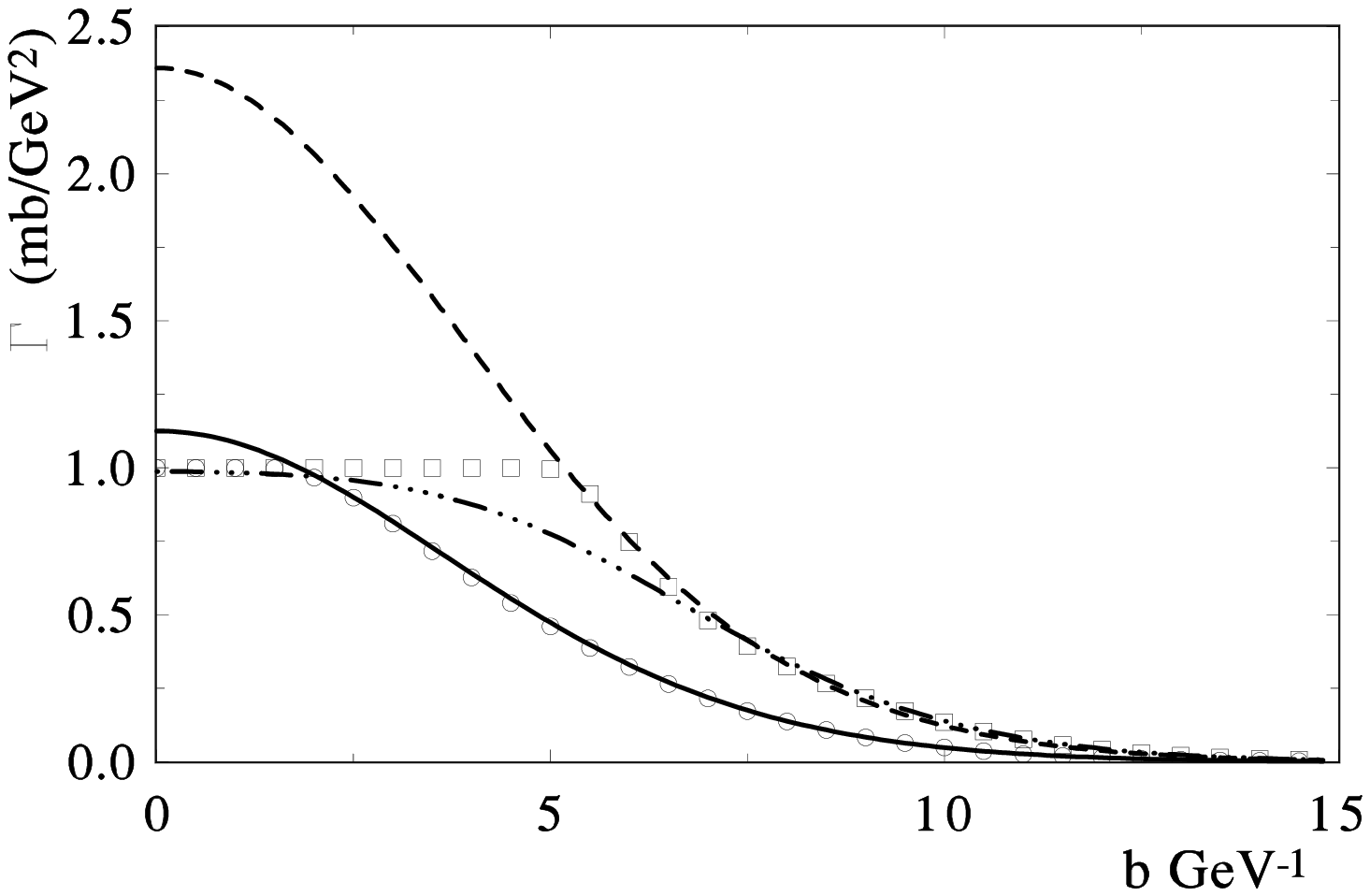}} \vspace{-0.5cm} \ec \vspace{-8mm}
\caption{The profile function of proton--proton scattering. (hard
line and circles -- at $\sqrt{s}=2$\,TeV without and with
saturation; dashed line and squares -- at $\sqrt{s}=14$\,TeV
without and with saturation; dash-dotted line -- the eikonal form
(\ref{eik-pf}) at $\sqrt{s}=14$\,TeV)} \label{Fig_1} \efg

Saturation of the profile function will control the behaviour of
$\sigma_{\mathrm{tot}}$ at super--high energies. Note that one
cannot simply cut the profile function sharply as this would lead
to a non-analytic amplitude, and to specific diffractive patterns
in the total cross section and in the slope of the differential
cross sections. Furthermore, we have to match at large impact
parameter the behaviour of the unsaturated profile function. We
have tried some specific matching patterns which softly
interpolate between both regimes. The interpolating functions give
unity in the large impact parameter region and force the profile
function to approach the saturation scale $b_s$ as a Gaussian.

If we take a single simple pole for the scattering amplitude, with
an exponential form factor, the radius of saturation, and its
dependence on energy, can be obtained analytically:
$$
R(s)^2\sim4\left[B\log\left(\frac{h}{2B}\right)+
\left(B+\log\frac{h}{2(B+\epsilon\log s}\right)\epsilon\log s+
\epsilon^2\log^2 s\right],
$$
where $B$ is the average slope at small $t$. Hence the total cross
section grows logarithmically at medium energies and grows like
$\log^2 s$ at very high energies.

\bfg[t] \vskip-15mm\bc \epsfysize=80mm
\centerline{\epsfbox{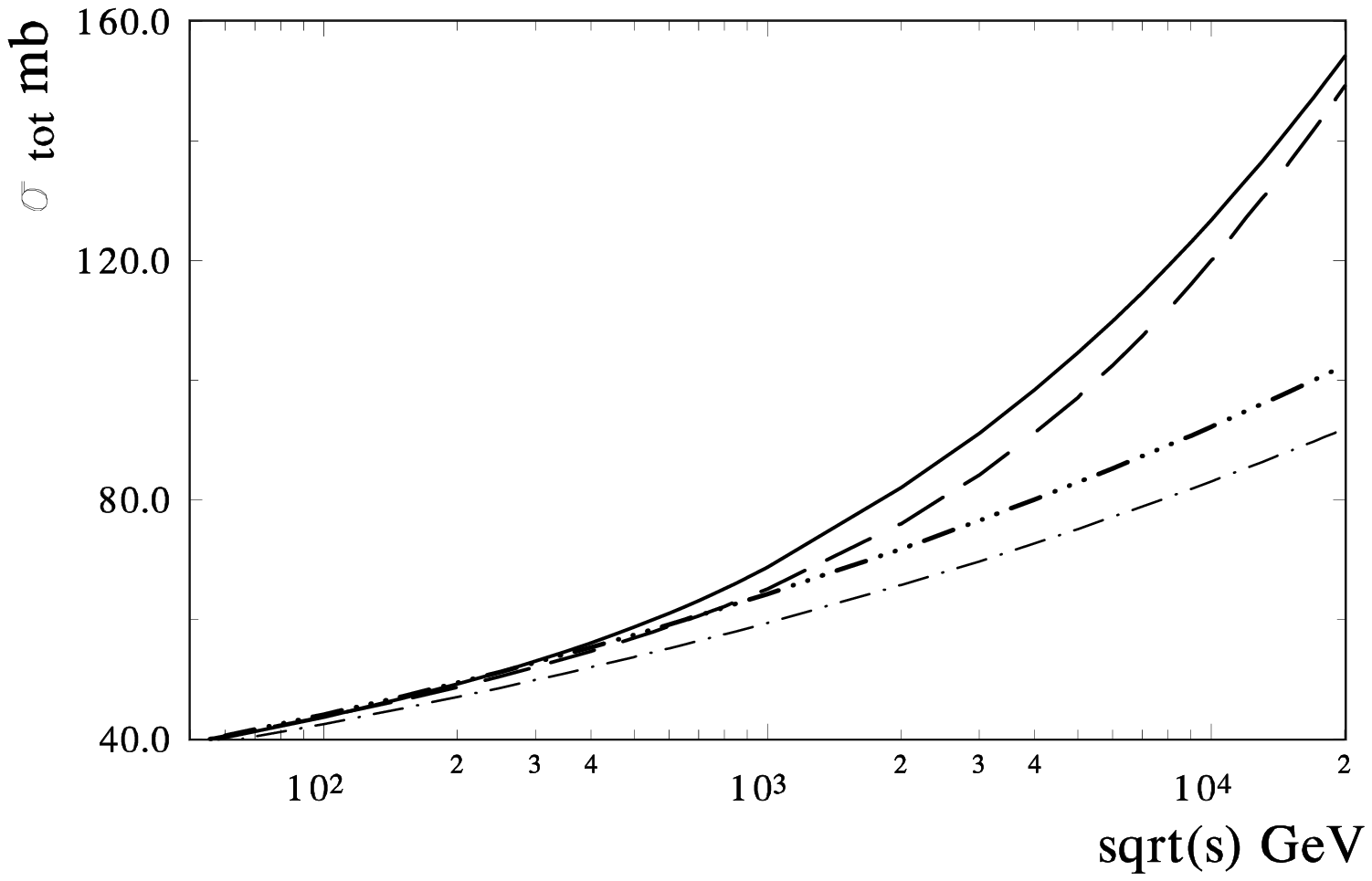}}
\vspace{-0.5cm} \ec \vspace{-8mm}%
\caption{$\sigma_{\mathrm{tot}}$ calculated in  different
approaches. (The hard line -- saturation regime; the long-dashed
line -- with the eikonal representation (\ref{eik-pf}); the
dash-dot-dotted line -- with soft pomeron and unitarized hard
pomeron \cite{mrt}; the dash-dotted line -- with  only a soft
pomeron)} \label{Fig_2}
\end{figure}

We show in Fig. 2 the possible behaviours of the total cross
section at very high energies, depending on the model and on the
unitarization scheme. We also show there the result of a simple
eikonalization, where we took \be T_{pp}(s,t=0)\,=\,2\int\D^2b\,
\left[1-\exp\bigl(h_{\mathrm{eik}}G(s,b)\bigr)\right].
\label{eik-pf} \ee and where $h_{\mathrm{eik}}=1.2$ was chosen so
that the values $\sigma_{\mathrm{tot}}$ determined  by
(\ref{eik-pf}) and by the saturation procedure are equal at
$\sqrt{s}=50$\,GeV, and $G(s,b)$ is the Fourier transform of $\Im T^0(s,t)$ 
given in (\ref{ampl}).

\section{ Non-linear equations}

The problem of the implementation of unitarity via saturation is
that the matching procedure seems arbitrary. Hence we considered a
different approach to saturate the amplitude. It is connected with
the non-linear saturation processes which have been considered in
a perturbative QCD context \cite{grib,mcler}. Such processes lead
to an infinite set of coupled evolution equations in energy for
the correlation functions of multiple Wilson lines
\cite{balitsky}. In the approximation where the correlation
functions for more than two Wilson lines factorize, the problem
reduces to the non-linear Balitsky--Kovchegov (BK) equation
\cite{balitsky,kovchegov}.

It is unclear how to extend these results to the non-perturbative
region, but one will probably obtain a similar equation. In fact
we found simple differential equations that reproduce either the
$U$--matrix or the eikonal representation. We can first consider
saturation equations of the general form
\begin{equation}
{\partial N(\xi,b)\over \partial\xi}=\S(N)\,,
\end{equation}
with $N$ the true (saturated) imaginary part of the amplitude. We shall
impose the following conditions:
\begin{itemize}
\item[(a)] $N\rightarrow 1$ as $s\rightarrow\infty$,
\item[(b)] $\partial N/\partial\xi\rightarrow 0$ as $s\rightarrow\infty$,
\item[(c)] $\S(N)$ has a Taylor expansion in $N$, and considering
the first term only gives the hard pomeron
$N_{\mathrm{bare}}=f(b)s^\Delta$. Similarly, we fix the
integration constant by demanding that the first term of the
expansion in $s^\Delta$ reduces to $N_{\mathrm{bare}}$.
\end{itemize}
Inspired by the BK results, we shall use the evolution variable
$\xi=\log s$. If we want to fulfil condition (c), then we need to
take $\S(N)=\Delta N+O(N^2)$. Conditions (a) and (b) then give
$\S(N)=\Delta(N-N^2)$ as the simplest saturating function.
The resulting equation
\begin{equation}
{\partial N\over \partial\log s}=\Delta (N-N^2)
\label{satu1}
\end{equation}
has the solution
\begin{equation}
N={f(b)s^\Delta\over f(b)s^\Delta + 1}\,. \label{solU}
\end{equation}
One can in fact go further: eq. (\ref{satu1}) has been written for
the imaginary part of the amplitude. If we want to generalize  it
to a complex amplitude, so that it reduces to (\ref{satu1}) when
the real part vanishes, we must take:
\begin{equation}
{\partial A\over \partial\log s}=\Delta(A+\I A^2)\,.
\label{satu1c}
\end{equation}
The solution of this is exactly the form (\ref{U-matrix}) obtained
in the $U$--matrix formalism, for $\Im U(s,b)=s^\Delta f(b)$.

Many other unitarization schemes are possible, depending on the
function ${\cal F}(N)$. We shall simply indicate here that the
eikonal scheme can be obtained as follows:
\begin{equation}
{\partial N\over\partial\log s}=
\Delta(1-N)\bigl(-\log(1-N)\bigr)\,.
\end{equation}

Other unitarization equations can be easily obtained via another
first--order equation. The idea here is that the saturation
variable is the imaginary part of the bare amplitude. One can then
write
\begin{equation}
{\partial N\over\partial N_{\mathrm{bare}}}= {\cal
F}'(N)\Rightarrow{\partial N\over \partial\log s}={\partial
N_{\mathrm{bare}}\over
\partial\log s}{\cal F}'(N)\,,
\end{equation}
with $N_{\mathrm{bare}}$ the unsaturated amplitude.

This will trivially obey the conditions (a)\,--\,(c) above, and
saturate at $N=1$. Choosing ${\cal F}'(N)=1-N$ gives the eikonal
solution
\begin{equation}
N(b,s)=1-\exp\bigl(-N_{\mathrm{bare}}(b,s)\bigr)\,, \label{sateik}
\end{equation}
whereas ${\cal F}'(N)=(1-N)^2$ leads to the $U$--matrix
representation (\ref{solU}).

We can come to the same results if we solve this equation via an
iteration procedure. For that let us take some model of the hadron
interaction in which the main hadron--hadron interaction is
created by the valence quarks surrounded by clouds of sea quarks.

Ref.~\cite{kovner1} proposed a picture in which the whole impact
parameter interval is divided into small and large distances: the
BFKL approximation works within a small radius $R_0$ and the BK
representation works for large impact parameters.

We examine two cases: the first includes only a short--range form
factor (as a simple Gaussian), and the second uses a short--range
form factor and a long--range one (in the form of a MacDonald
function).

So, we divide the whole energy interval  in small pieces inversely
proportional to $s^{2\Delta}$. We then obtain the step in $s$ \be
s_0=s^{\Delta}/s^{2 \Delta}=1/ s^{\Delta} \ee and the number of
such pieces will be \be n=\mathrm{int}\,(s^{2\Delta})\,, \ee where
the function $\mathrm{int}$ chooses the nearest natural number.

$N(s,b)$ after this small energy interval will be \be
N_0(s_0,b)=s_0f(b)=f(b)/s^{\Delta}\,. \ee We calculate the
derivative $\delta$ on this small interval and increase $N(s_0,b)$
by $\delta N_0(s_0,b)$: \be\ba{rcl}
\delta_i&=&1-N_i\,,\\[2pt]
N_{i+1}&=&N_{i}+\delta\,N_0(s_0,b)\,.\ea\ee We then iterate the
above procedure for $N_i=N(s_i,b)$ until we get to the end of the
interval. It is clearly understood that, if the energy is
sufficiently high, we obtain for some iteration $N_k=1$ and
$\delta=0$, so we reach the saturation bound at some impact
parameter. Of course this energy will depend on the form of
$f(b)$.

\bfg[b] \vskip-15mm\bc \epsfysize=80mm
\centerline{\epsfbox{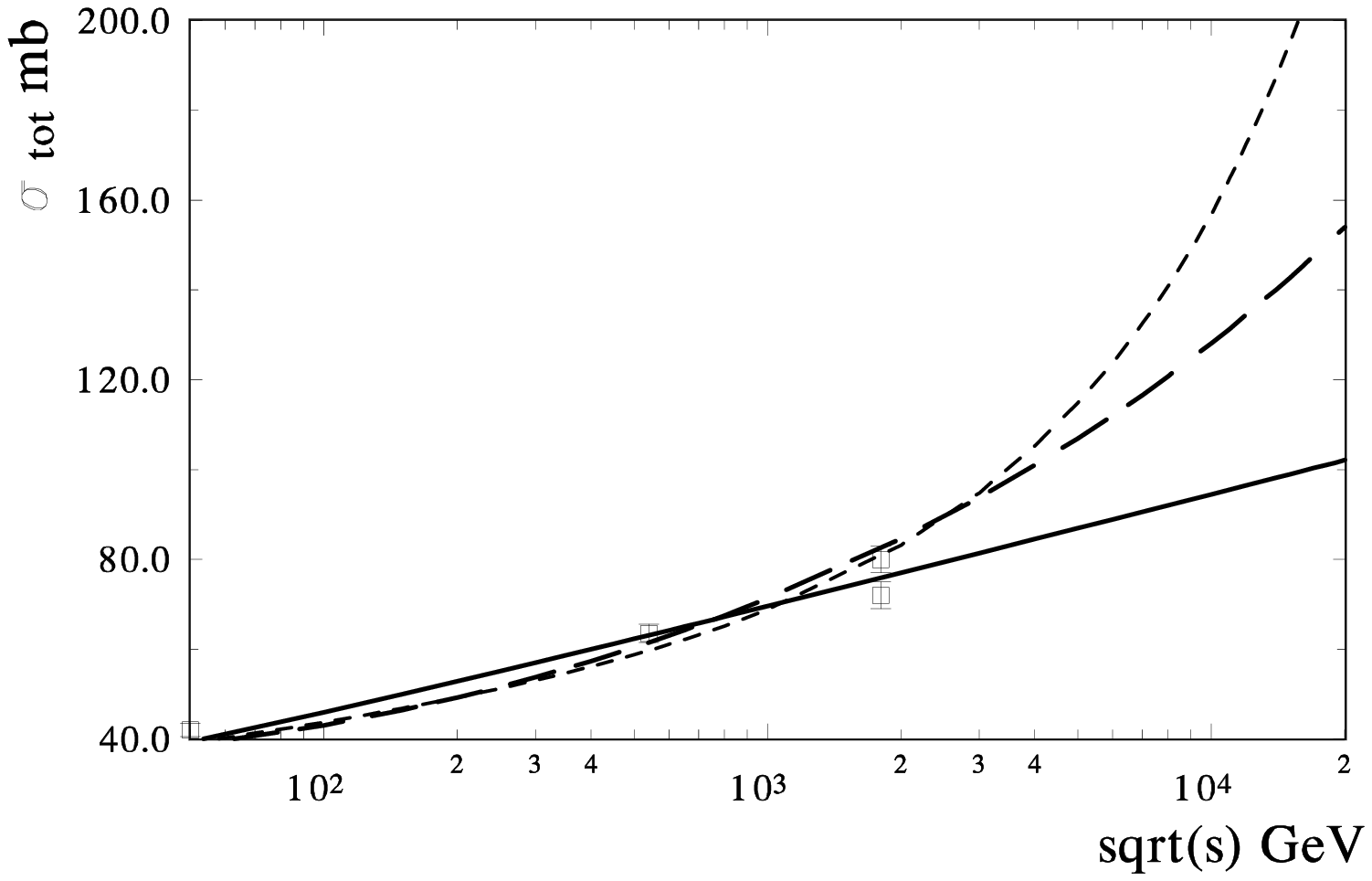}} \vspace{-12mm} \ec
\caption{$\sigma_{\mathrm{tot}}$ calculated in different
approaches. (The hard line -- $f(b)$ in Gaussian form; the
long-dashed line -- with the contribution of the large distances;
the dashed line -- in the soft\,+\,hard pomeron model)}
\label{Fig_3} \efg

In this case \be T(s,t=0)\,=\,\int_0^{\infty}b\,\D b\,N(s,b) \ee
and \be \sigma_{\mathrm{tot}}(s)=4\pi\ \Im T(s,t=0)\,. \ee The
results of our calculations are shown in Fig.~3. We can see that
energy dependence of the total cross section is a simple
logarithm. So, in this case, $\sigma_{\mathrm{tot}}(s)$ does not
saturate the Froissart unitarity bound and coincides with the
eikonal solution with a Gaussian eikonal.

Note that we obtain in this case the right energy dependence not
only for $\sigma_{\mathrm{tot}}(s)$,  but also for the
differential cross sections. Of course, this picture is only
qualitative. For a quantitative description of the different
features of the diffraction processes, we presumably need to take
into account many different effects and a more complicated
structure for the non-linear equation.

We have shown that the most usual unitarization schemes could be
recast into differential equations which are reminiscent of
saturation equations \cite{balitsky,kovchegov}. Such an approach
can be used to build new unitarization schemes and may also shed
some light on the physical processes underlying the saturation
regime.
\subsection*{Acknowledgement:}We are grateful to S.M. Troshin for
reading and correcting our original manuscript.

\end{document}